\documentclass{jot}

\usepackage{tikz}
\usetikzlibrary{shapes.geometric, arrows.meta, positioning}
\usepackage{amssymb}
\usepackage{booktabs}
\usepackage{bm}
\usepackage{comment}

\title{Modeling and Link Budget Feasibility Analysis of Secure LoRa-Based Peer-to-Peer Communication for Short-Range Tactical Networks}


\Instytucja{UPES}
{School of Computer Science, UPES, Dehradun, India}
{}

\Instytucja{NIT}
{Department of Electronics and Communication Engineering, NIT Jamshedpur, Jamshedpur, India}
{}

\Author{}{Ayush Kumar Agrawal}
{ayush6295@gmail.com}
{}
{UPES}
{}
{}
{}
{School of Computer Science}
{}

\Author{}{Soumendu Das}
{soumendu.das13@gmail.com}
{}
{UPES}
{}
{}
{}
{School of Computer Science}
{}

\Author{}{Rashmi Keshri}
{rashmikeshri28@gmail.com}
{}
{UPES}
{}
{}
{}
{School of Computer Science}
{}

\Author{}{Saptaparna De}
{desaptaparna@gmail.com}
{}
{NIT}
{}
{}
{}
{Department of Electronics and Communication Engineering}
{}

\Author{}{Jayendra Kumar}
{jkumar.ece@nitjsr.ac.in}
{}
{NIT}
{}
{}
{}
{Department of Electronics and Communication Engineering}
{}

\begin{document}
\maketitle

\begin{Abstract}
Short-range reliable and secure communication is a major priority in the tactical, military and disaster response settings where the traditional communication infrastructure is either off-line or prone to interception. Current VHF/UHF radios and software-defined radios are popular but large-sized devices and require lots of power, making them not suitable to be used as lightweight wearable devices with seamless hand-free use. In this paper, the design and theoretical framework of a miniature, LoRa based encrypted intercommunication device that can be used in secure field communication over a range of 1-1.5km and under line-of-sight conditions is provided. The suggested system consists of a voice-activated acquisition block, digital audio compression, an embedded microcontroller processor, and AES-128 encryption followed by a low-power transmission via the LoRa protocol. Through the ability of chirp spread spectrum modulation to utilize the long-range and low-energy properties, the system is guaranteed reliable communications coupled with low power consumption and low electromagnetic footprint. The theoretical analysis of the proposed communication range is justified using a link-budget that justifies the practicability of the communication range in the real propagation conditions. This architecture focuses on infrastructural agnosticism, peer-to-peer security as well as wearable ergonomics. The given scheme shows the possibilities of LoRa technology in the scope of other traditional IoT telemetry, and it can be further extended to include secure tactical voice communication platforms.
\end{Abstract}

\section{Introduction}\label{s:1}
Effective communications in dynamic and hostile field environments are an essential need regarding operations of mission critical issues. The operations of military patrol, border surveillance, disaster response, and emergency rescue operators are often provided in the area where the standard communication infrastructure, including the cellular system or satellite connection, is either absent, overloaded, or exposed to interference and compromise. The communication systems in these cases have to meet the strict criteria such as low latency, high reliability, low detectability and high cryptographic security.

Tactical communication has traditionally been performed by the use of radio systems of both VHF and UHF. Although these systems are able to offer moderate range and high robustness, they have a number of practical limitations. Their hardware design is generally huge and consumes energy causing lack of mobile capability and restricting of wearability. Besides, a large number of old systems are based on the push-to-talk systems and external antenna systems, which may impede the smooth hands-free communication during the active operations. Security wise even modern variants have added encryption modules but secure key exchange and end-to-end confidentiality is not always efficiently designed on the embedded level. Also, the ongoing high-power transmission enhances electromagnetic visibility which may reveal operational presence.

The recent development of low-power wide-area network technologies, especially the LoRa technology based on the chirp spread spectrum modulation, offers a new paradigm of short-range tactical communications. LoRa is said to have a high receiver sensitivity, a large link budget, and a low energy consumption which makes it theoretically viable in small wearable systems. Nevertheless, it has been mostly used in low-data-rate telemetry and sensor networks as opposed to real-time encrypted voice transmission. In an effort to close this gap, this paper presents a wearable LoRa-based safe voice intercommunication platform intended to be deployed in a short-range in peer-to-peer mode of application \cite{ref1}. The system combines voice activation features, digital audio compression, integrated microcontroller processing and AES-based encryption before the LoRa transmission. The proposed architecture will deliver infrastructure independence, footprint reduction and increased operation security, but support secure exchange of packets and voice and guarantee reliable exchange of voice over a 1-1.5 km line-of-sight range.

\subsection{Novelty and Contributions}\label{s:1.1}
Although LoRa technology has been extensively studied for telemetry-oriented Internet of Things deployments, its use for secure real-time voice communication in tactical peer-to-peer environments remains largely unexplored. Existing LoRaWAN systems primarily focus on low-data-rate sensing applications and rely on centralized gateway architectures, which introduce infrastructure dependency and additional latency.

The novelty of this work lies in extending LoRa communication beyond conventional telemetry applications toward secure tactical voice communication. Unlike traditional LoRaWAN architectures, the proposed framework operates in a decentralized peer-to-peer configuration and integrates encryption directly within the embedded firmware layer before packet transmission. Furthermore, the system combines voice-activated acquisition, low-bitrate audio compression, AES-128 encryption, and chirp spread spectrum modulation to enable secure voice exchange over low-power LoRa links.

The main contributions of this work are summarized as follows:
\begin{itemize}
\item Design of a decentralized LoRa-based secure voice communication framework for short-range tactical deployments.
\item Integration of firmware-level AES encryption with voice packetization for end-to-end confidentiality.
\item Analytical evaluation of communication feasibility through link budget and latency modeling.
\item Comparative analysis with existing communication technologies including LoRaWAN, WiFi ad hoc networks, and conventional VHF/UHF radios.
\end{itemize}

\section{Theoretical Comparative Analysis}\label{s:2}
The development of the secure short-range ad hoc communications needs the multidimensional evaluation framework wherein the architectural decentralization, level of encryption integration, latency, energy efficiency, and independence of the infrastructure have to be taken into consideration. Although the LoRa based systems have been studied extensively in the context of telemetry based Internet of Things (IoT) deployment, little has been done to extend the system to support real time secure voice exchange in tactical peer-to-peer networks.

The majority of LoRaWAN based architecture is based on the star-of-stars topology where the end nodes interact with centralized gateways and network servers. By its very nature such designs bring about infrastructure dependency and the added latency caused by the processing of packet requests by a cloud service. AES encryption is also supported in LoRaWAN, but it is often implemented on the network layer as opposed to being closely integrated in embedded firmware to support real time encrypted voice payload communications.

Contrary to this, IEEE 802.11 ad hoc networks have direct node-to-node communication and low transmission latency. However, they have higher bandwidth, carrier frequencies, which means that they consume a lot of energy, and have a low range of communication when constrained by portable power sources. Likewise, the traditional VHF/UHF tactical radios support medium-range voice communications but are usually not adaptively duty-cycled, with embedded encryption at the microcontroller level, and can be fabricated into a small wearable form factor.

To explicitly point out these differences, Tab.~\tref{1} gives an architectural-level comparison in terms of technology type, application orientation, depth of encryption implementation, and dependency of infrastructure.

\begin{table*}
\Caption{Architectural-Level Comparison of Communication Approaches.}\tlabel{1}
\begin{tabular}{|C{2.8cm}|C{1.8cm}|C{2.2cm}|C{1.5cm}|C{1.8cm}|C{2.1cm}|}
\hline
\textbf{Method} & \textbf{Technology} & \textbf{Application Focus} & \textbf{Voice Support} & \textbf{Encryption} & \textbf{Infrastructure Dependency} \\\hline
LoRa Telemetry Systems & LoRa & IoT Telemetry & No & Network-Level AES & Gateway Required \\\hline
LoRaWAN Architectures & LoRaWAN & IoT Scalability & No & AES (Network Layer) & Centralized Server \\\hline
WiFi Ad Hoc Networks & IEEE 802.11 & Data Communication & Yes & WPA2/WPA3 & Independent but High Power \\\hline
Conventional VHF/UHF Radios & Analog/Digital RF & Tactical Voice & Yes & Limited / External & Independent \\\hline
Proposed System & LoRa (CSS) & Secure Tactical Voice & Yes & AES-128 (Firmware-Level) & Infrastructure-Free Peer-to-Peer \\\hline
\end{tabular}
\end{table*}

In addition to architectural factors, quantitative characteristics of performance define whether ad hoc tactical communication systems can be operational or not. System effectiveness is controlled by three major parameters:
\begin{enumerate}
\item Communication Range: This is dependent on link budget, receiver sensitivity and modulation robustness.
\item End-to-End Latency: This is a combination of encoding delay, encryption overhead, packet airtime and decoding time.
\item Energy Consumption: Controlled by transmission duty cycle, RF power level and sleep scheduling efficiency.
\end{enumerate}

Conventional LoRa telemetry implementations favor a greater range at the expense of high airtime and greater spreading factors which gives latency values that cannot support conversational interaction. On the other hand, ad hoc systems based on WiFi are characterized by low latency although they require constant high power consumption because of carrier sensing and contention. Traditional tactical radios are more powered, thus costing more in battery drainage and electromagnetic emissions.

Tab.~\tref{2} displays the main operational characteristics of the prototypical communication technologies to obtain a quantitative positioning of the proposed system within this design space.
\begin{table}
\Caption{Performance Comparison of Communication Technologies.}\tlabel{2}
\begin{tabular}{|C{2.4cm}|C{1.3cm}|C{1.5cm}|C{1.5cm}|}
\hline
\textbf{Technology} & \textbf{Range} & \textbf{Latency} & \textbf{Power Consumption} \\\hline
LoRa Telemetry Systems & 2--15 km & High (Seconds) & Very Low \\\hline
WiFi Ad Hoc Networks & 100--300 m & $<$50 ms & High \\\hline
VHF/UHF Radios & 1--5 km & Low & High \\\hline
Proposed System & 1--1.5 km & $\leq$ 300 ms & Low \\\hline
\end{tabular}
\end{table}

The comparison of performance shows that the proposed system has a balanced trade-off in the range, latency, and energy consumption. Although LoRa telemetry networks have multi-kilometer ranges, their spreading factors can cause air time delays that are too high to permit real time voice communications. Ad hoc WiFi has lower delay but low operational performance when deployed on wearable sensors powered by batteries.

The proposed architecture is able to maintain communication over a 1-1.5 km line-of-sight distance and conserve conversational latency at 300 ms. This is done by the use of optimal spreading factor choice and controlled packet size. Further, voice-activated transmission ensures that average duty cycle is also cut to a large extent, enhancing battery sustainability without compromises in responsiveness.

Theoretically, the proposed system lies in the middle of the gap between the telemetry models of the LPWAN and the models of high-throughput ad hoc networks by being efficient enough to bridge the energy concerns and the latency of the secure voice exchange with the decentralized tactical setting.

\section{System Architecture}\label{s:3}
\subsection{Overall Communication Model}\label{s:3.1}
The proposed architecture is based on a modular, layered architecture that is expected to provide secure, low-latency and energy-efficient voice communications in the field. The general communication topology is shown in Fig.~\fref{1}.
\begin{figure}
\ffig[width=83mm]{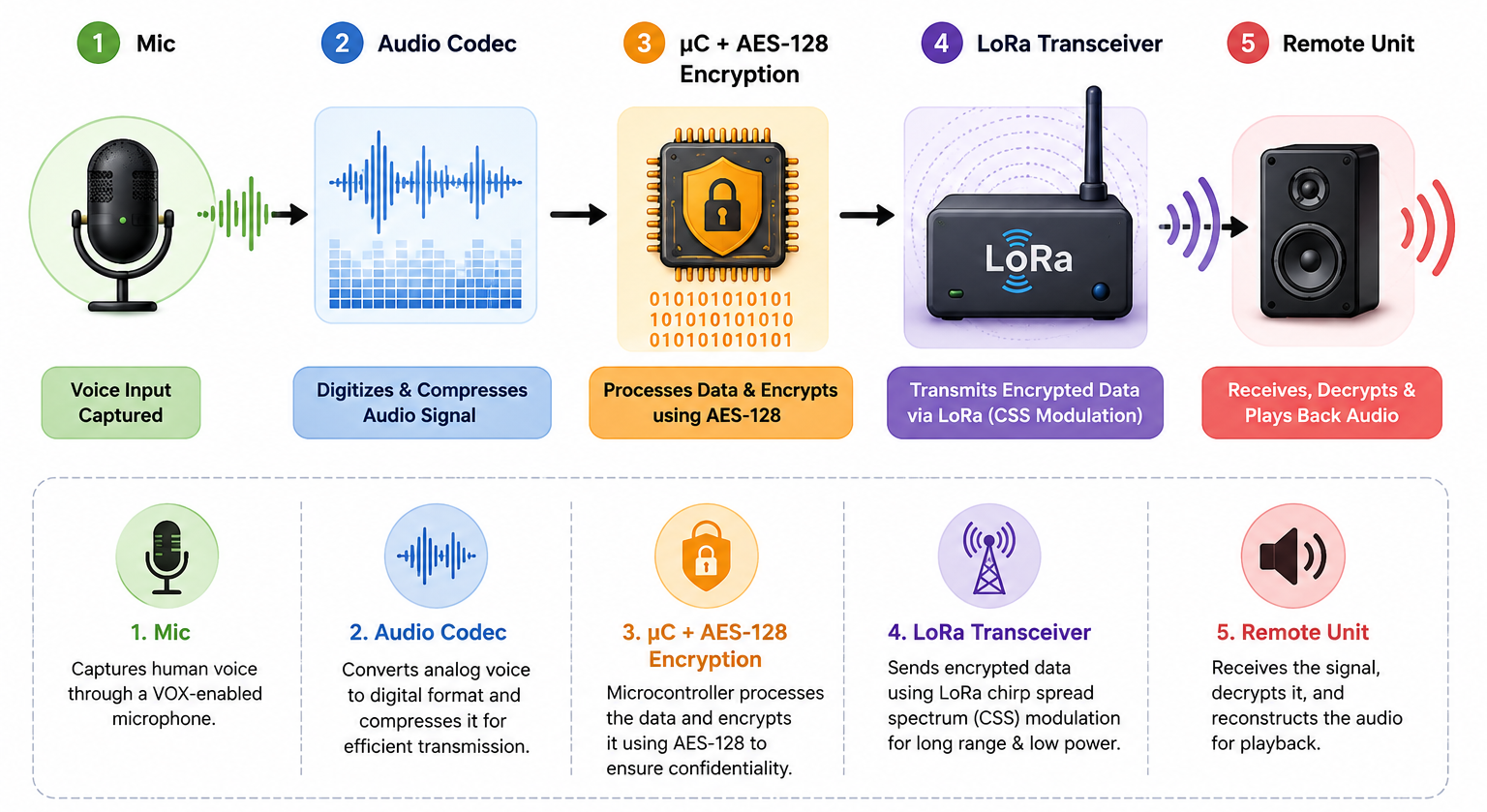}[0mm]
\Caption{Overall Communication Flow.}\flabel{1}
\end{figure}
The communication pipeline is built up by a series of stages of signal transformation:
\begin{center}
VOX Microphone $\rightarrow$ Audio Codec $\rightarrow$ MCU Processing $\rightarrow$ AES Encryption $\rightarrow$ LoRa Transmission $\rightarrow$ Remote Decryption and Playback
\end{center}

The analog voice data is sampled in the input stage by a VOX enabled microphone module. The voice operated exchange (VOX) mechanism is triggered only when the signal amplitude is above a certain threshold which helps to reduce idle power consumption and suppress unintended RF emissions, as well as any unintended RF emission caused by the power consumption \cite{ref2}.

The audio codec encodes the analog signal into a compressed stream at a low bit-rate encoded into digital form, and this can be transmitted by LoRa within its bandwidth limitations. The microcontroller unit (MCU) does buffering, frame packeting, cyclic redundancy check (CRC) addition, as well as timing synchronisation \cite{ref3}. Before transmission, AES-128 encryption is done to guarantee confidentiality and integrity. The coded message is then interspersed with the chirp spread spectrum (CSS) method of LoRa, and sent over the ISM band.

The same work is carried out at the receiving end which involves demodulation, decryption, decoding and reconstruction of the audio signal \cite{ref4}. The same is done at the receiving end which includes demodulation, decryption, decoding and reconstruction of the audio signal.

\subsection{Peer-to-Peer Subsystem Architecture}\label{s:3.2}
The subsystem architecture as shown in Fig.~\fref{2} is a mirrored subsystem architecture, which depicts two wearable field units in symmetric peer-to-peer mode.
\begin{figure}
\ffig[width=83mm]{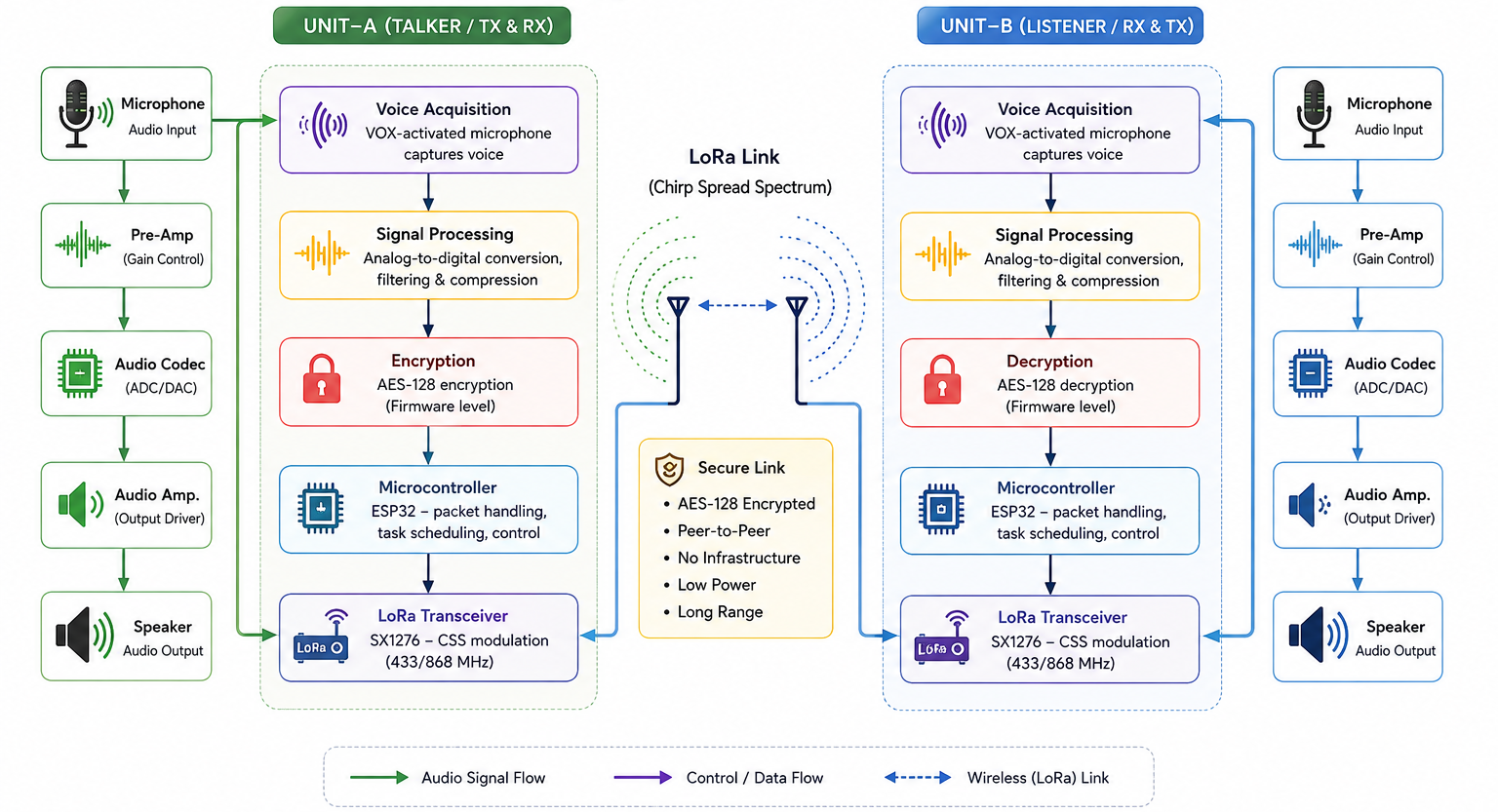}[0mm]
\Caption{Full-Duplex Subsystem Architecture.}\flabel{2}
\end{figure}

The functional layer of each subsystem consists of the following:

\textbf{1. Voice Acquisition Layer:}
The VOX enabled microphone constantly checks the ambient sound and when the system detects speech, it is activated. This lowers the duty cycle and increases battery life.

\textbf{2. Signal Processing Layer:}
Audio codec does analog to digital conversion and compression. Since LoRa has the low data rate, the compression ratio is adjusted to trade intelligibility against transmission efficiency.

\textbf{3. Control and Security Layer:}
Task scheduling, packet segmentation, encryption and error handling is done with an ESP32 microcontroller. The hardware acceleration is used to ensure that the implementation of AES-128 encryption is used to reduce the number of computations.

\textbf{4. Communication Layer:}
The SX1276 LoRa transceiver is a chirp spread spectrum modulation of encrypted packets. LoRa has a very high receiver sensitivity (as high as -148 dBm) and allows long distances to be covered with a small transmit power.

\textbf{5. Audio Reconstruction Layer:}
Decrypted digital audio is decoded on reception and reproduced using a small speaker module.

The architecture is in a half-duplex mode where the process of transmission and reception are alternated \cite{ref5}. The architecture is half-duplex mode that is, the process of transmissions and receipts takes place in turns. This minimizes the chances of collisions, eases synchronization and minimizes the average power consumption.

\subsection{Hardware Layout and Physical Design}\label{s:3.3}
The Fig.~\fref{3} shows the physical set up of the device.
\begin{figure}
\ffig[width=83mm]{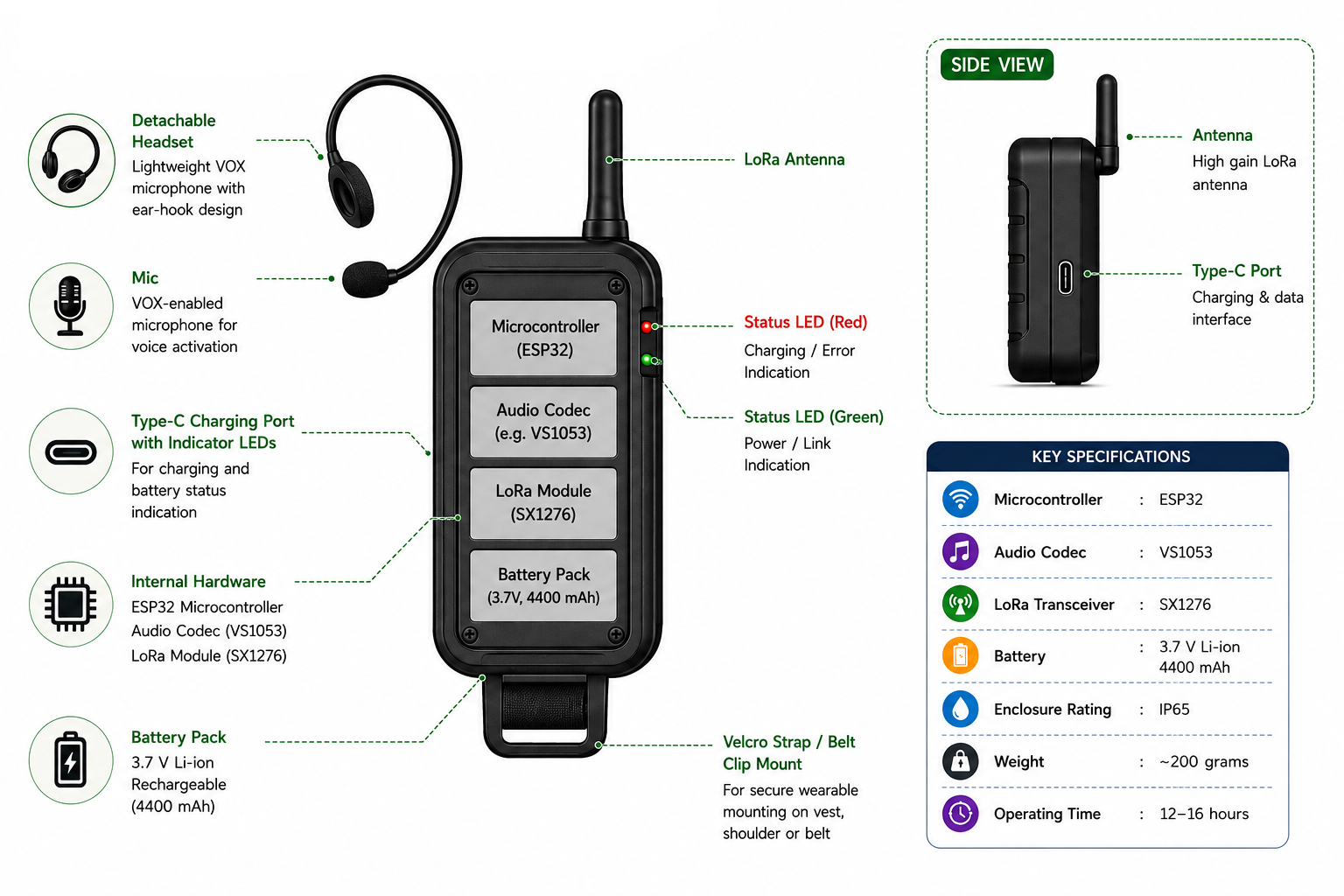}[0mm]
\Caption{Hardware Layout of Proposed Device.}\flabel{3}
\end{figure}
The hardware design gives more emphasis to field adaptability, ruggedization and portability. ABS material is used to make the enclosure with IP65 rating material to resist dust and water ingress. The system stack includes internal hardware which comprises:
\begin{itemize}
\item ESP32 microcontroller module
\item VS1053 or similar low-power audio codec
\item SX1276 transceiver LoRa module
\item 3.7V Li-ion rechargeable battery
\item built-in charging and power regulation circuitry.
\end{itemize}

\begin{figure}
\ffig[width=83mm]{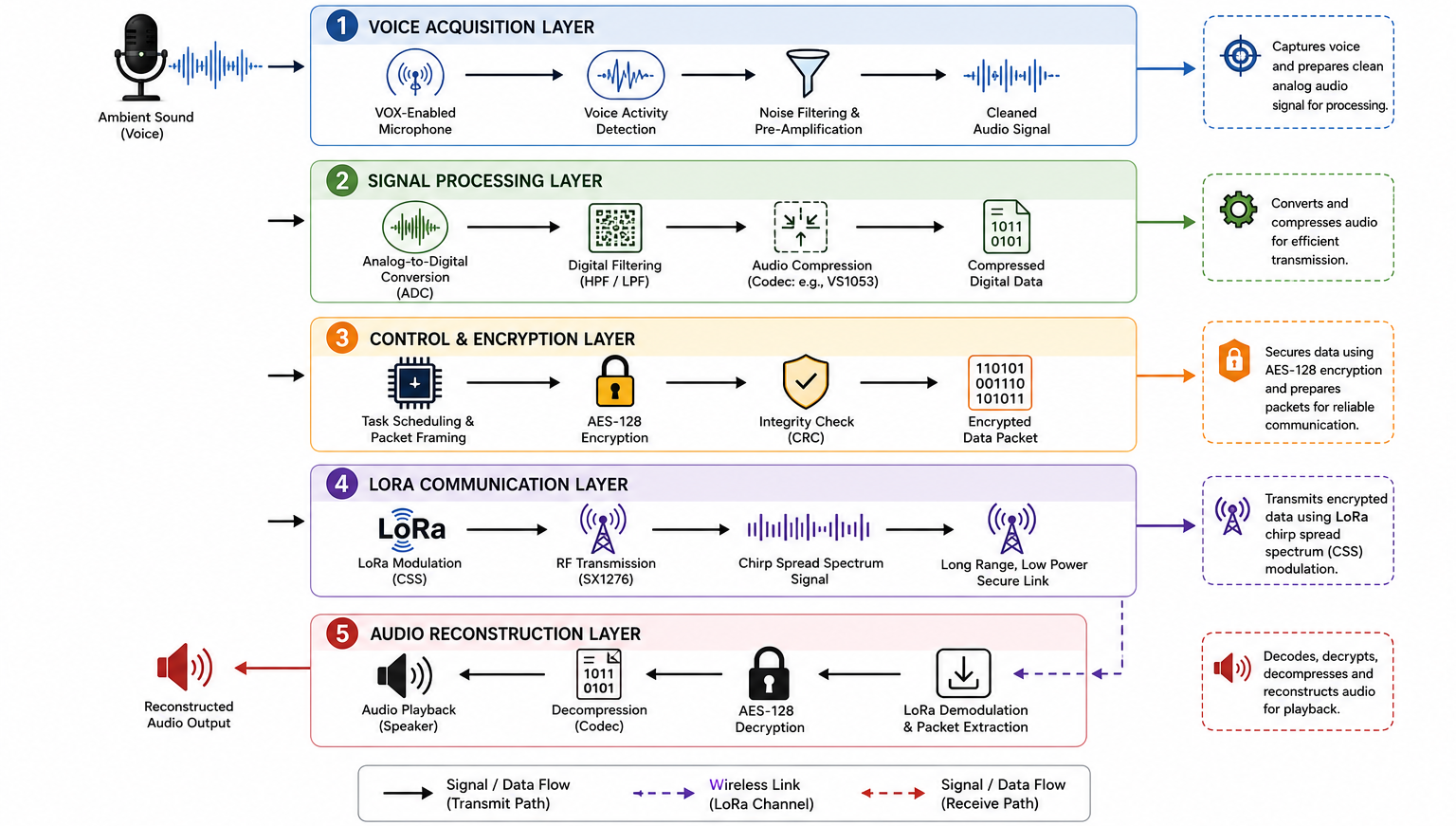}[0mm]
\Caption{Layered Functional Architecture of the Proposed System.}\flabel{4}
\end{figure}

The removable lightweight headset makes its use in lengthy missions ergonomic in the headset \cite{ref6}. In the headset, the detachable lightweight headset makes its use economical throughout lengthy missions. The antenna interface is mounted externally so that it makes the maximum radiation efficiency. The weight of the device is kept under 200 grams, to make it comfortable in terms of the way it can be worn in its deployment \cite{ref7}.
Strategies of power optimization involve:
\begin{itemize}
\item VOX-triggered activation
\item MCU deep sleep during idle state
\item Adaptive transmission power control
\item Efficient packet scheduling
\end{itemize}
Continuous operation of 12--16 hours is ensured under typical field usage conditions.

\section{Technical Specifications}\label{s:4}
\subsection{Communication Parameters}\label{s:4.1}
The proposed communication platform works with the band of Industrial, Scientific and Medical, 433 MHz and 868 MHz in their operation \cite{ref8}. The reason why these bands are selected is due to their good propagation properties, regulatory accessibility and conformity to LoRa chirp spread spectrum modulation \cite{ref9}.
The key working parameters are enumerated below:
\begin{itemize}
\item \textbf{Frequency Band:} 433 / 868 MHz (ISM band)
\item \textbf{Operational Range:} 1--1.5 km (Line-of-Sight)
\item \textbf{Latency:} $\leq 300$ ms end-to-end
\item \textbf{Encryption:} AES-128 symmetric key encryption
\item \textbf{Mode of Operation:} Half-duplex
\item \textbf{Transmit Power:} 14--20 dBm
\item \textbf{Receiver Sensitivity:} up to $-148$ dBm (depending on spreading factor)
\item \textbf{Battery Backup:} 12--16 hours continuous operation
\end{itemize}

LoRa modulation allows high sensitivity of receivers in chirp spread spectrum (CSS) methods \cite{ref10}. An improved spreading factor contributes to link robustness at the cost of decreased data rate, enabling a viable trade-off between range and latency \cite{ref11}.

It has been chosen that half-duplex communication would be used to decrease the probability of collisions and decrease the average current consumption \cite{ref12}. AES-128 is very secure and has minimal computational cost and thus it is applicable in tactical communication environment \cite{ref13}.

Half-duplex communication is adopted to reduce collision probability and minimize average current consumption \cite{ref12}. AES-128 encryption ensures computational efficiency while maintaining strong cryptographic security suitable for tactical communication \cite{ref13}.
The End-to-end latency may be stated as a total of all the latencies as follows:
\begin{equation}\elabel{e1}
T_{\text{total}} = T_{\text{encoding}}
+ T_{\text{encryption}}
+ T_{\text{packetization}}
+ T_{\text{airtime}}
+ T_{\text{decoding}}
\end{equation}
The system maintains below 300 ms latency with optimised compression and moderate spreading factors (e.g. SF7- SF9) and is therefore acceptable in conversational voice communication.

\subsection{Link Budget Analysis}\label{s:4.2}
From a physical-layer perspective, LoRa communication relies on chirp spread spectrum (CSS) modulation, which provides high receiver sensitivity and strong resistance to interference and multipath fading. Unlike conventional narrowband modulation schemes, CSS spreads the transmitted signal over a wider bandwidth using frequency-modulated chirps. This technique improves the signal-to-noise ratio at the receiver through processing gain and enables reliable reception even under weak signal conditions.

The achievable communication range in LoRa systems is primarily governed by the link budget, which represents the balance between transmitted power, propagation losses, antenna gains, and receiver sensitivity. Due to its high sensitivity (often below -130 dBm), LoRa can maintain connectivity at relatively low transmit power levels, making it suitable for energy-constrained wearable communication devices.

In tactical environments where communication nodes may operate without infrastructure support, evaluating the theoretical link budget becomes essential for validating the feasibility of the proposed communication range. The communication range is based on the theoretical feasibility which is confirmed through an ordinary link budget formulation. The signal power that is received is determined as:
\begin{equation}\elabel{e2}
P_r = P_t + G_t + G_r - L_p - L_s
\end{equation}
where:
\begin{itemize}
\item $P_t$ = Transmit power (dBm)
\item $G_t$ = Transmitter antenna gain (dBi)
\item $G_r$ = Receiver antenna gain (dBi)
\item $L_p$ = Free space path loss (dB)
\item $L_s$ = System losses (cable, impedance mismatch, fading margin)
\end{itemize}
The free-space path loss (FSPL) is estimated using the well-known relation
\begin{equation}\elabel{e3}
L_{p} = 20\log_{10}(d) + 20\log_{10}(f) + 32.44
\end{equation}
where $d$ denotes the link distance in kilometers and $f$ represents the carrier frequency in MHz.
For the considered communication scenario,
\begin{equation}\nonumber
d = 1.5 \; \text{km}, \qquad f = 868 \; \text{MHz}.
\end{equation}
Substituting these values into Eq.~\eref{e3} gives
\begin{align}
L_{p}
&= 20\log_{10}(1.5) + 20\log_{10}(868) + 32.44 \\
&= 3.52 + 58.77 + 32.44 \\
&\approx 94.73 \; \text{dB}.
\end{align}
The received power is obtained from the standard link budget expression
\begin{equation}\elabel{e4}
P_{r} = P_{t} + G_{t} + G_{r} - L_{p} - L_{s}
\end{equation}
where $P_t$ is the transmit power, $G_t$ and $G_r$ denote the transmitter and receiver antenna gains, and $L_s$ accounts for miscellaneous system losses.
The following values are assumed:
\begin{equation}\nonumber
P_t = 17 \; \text{dBm}, \quad
G_t = 2 \; \text{dBi}, \quad
G_r = 2 \; \text{dBi}, \quad
L_s = 5 \; \text{dB}.
\end{equation}
Using Eq.~\eref{e4}, the received power becomes
\begin{align}
P_r
&= 17 + 2 + 2 - 94.73 - 5 \\
&\approx -78.73 \; \text{dBm}.
\end{align}

The link margin can therefore be written as
\begin{equation}\elabel{e5}
\text{Link Margin} = P_r - P_{\text{sensitivity}}.
\end{equation}
LoRa receivers typically achieve sensitivities in the range of approximately
$-130$ dBm to $-148$ dBm depending on the spreading factor.
Even under a conservative assumption of
\begin{equation}\nonumber
P_{\text{sensitivity}} = -120 \; \text{dBm},
\end{equation}
the resulting link margin is
\begin{align}
\text{Link Margin}
&= -78.73 - (-120) \\
&\approx 41 \; \text{dB}.
\end{align}
This large connection margin verifies good communication over the range of 1-1.5 km under line of sight, and can tolerate moderate environmental degradation and interference.

\subsection{Power Consumption Analysis}\label{s:4.3}
The duty cycling and VOX-controlled transmission impact the average power consumption as folllows:
\begin{equation}\elabel{e6}
P_{avg} = D \cdot P_{tx} + (1-D) \cdot P_{sleep}
\end{equation}
where $D$ is transmission duty cycle. As Voice transmission is intermittent, and, therefore, the value of $D$ is comparably small, which contributes to a much longer battery life. Its optimized firmware scheduling of the firmware sleep and adaptive transmit power control have a continuous operational readiness of 12 to 16 hours of long duration \cite{ref14}.

In order to quantitatively evaluate the feasibility of the proposed LoRa communication link, the key parameters used in the link budget calculation are summarized in Tab.~\tref{3}. These parameters represent typical operational values for low-power LoRa communication systems operating in the ISM band. The transmit power and antenna gains correspond to commonly used LoRa transceiver modules such as the SX1276, while the receiver sensitivity reflects the high sensitivity achievable through chirp spread spectrum modulation \cite{ref1,ref4}. The assumed system losses include cable attenuation, impedance mismatch, and environmental fading margins typically considered in practical wireless communication deployments \cite{ref6}. These parameters collectively allow the estimation of received signal strength and link margin, thereby validating the achievable communication distance under line-of-sight conditions.
\begin{table}
\Caption{Parameters Used in Link Budget Analysis.}\tlabel{3}
\begin{tabular}{|C{3.4cm}|C{3.2cm}|}
\hline
Parameter & Value \\\hline
Transmit Power ($P_t$) & 17 dBm \\\hline
Tx Antenna Gain ($G_t$) & 2 dBi \\\hline
Rx Antenna Gain ($G_r$) & 2 dBi \\\hline
Carrier Frequency & 868 MHz \\\hline
Distance & 1.5 km \\\hline
System Loss ($L_s$) & 5 dB \\\hline
Receiver Sensitivity & -120 dBm \\\hline
\end{tabular}
\end{table}

\section{Development Methodology}\label{s:5}
A well structured hardware and software co-design procedure was followed in the development work of the proposed LoRa based secure intercommunication device. This process was carefully structured around sequential but iterative phases, and as such ensured reliability, scalability and field-readiness. The overall development flow is shown in Fig.~\fref{5}.
\begin{figure}
\ffig[width=83mm]{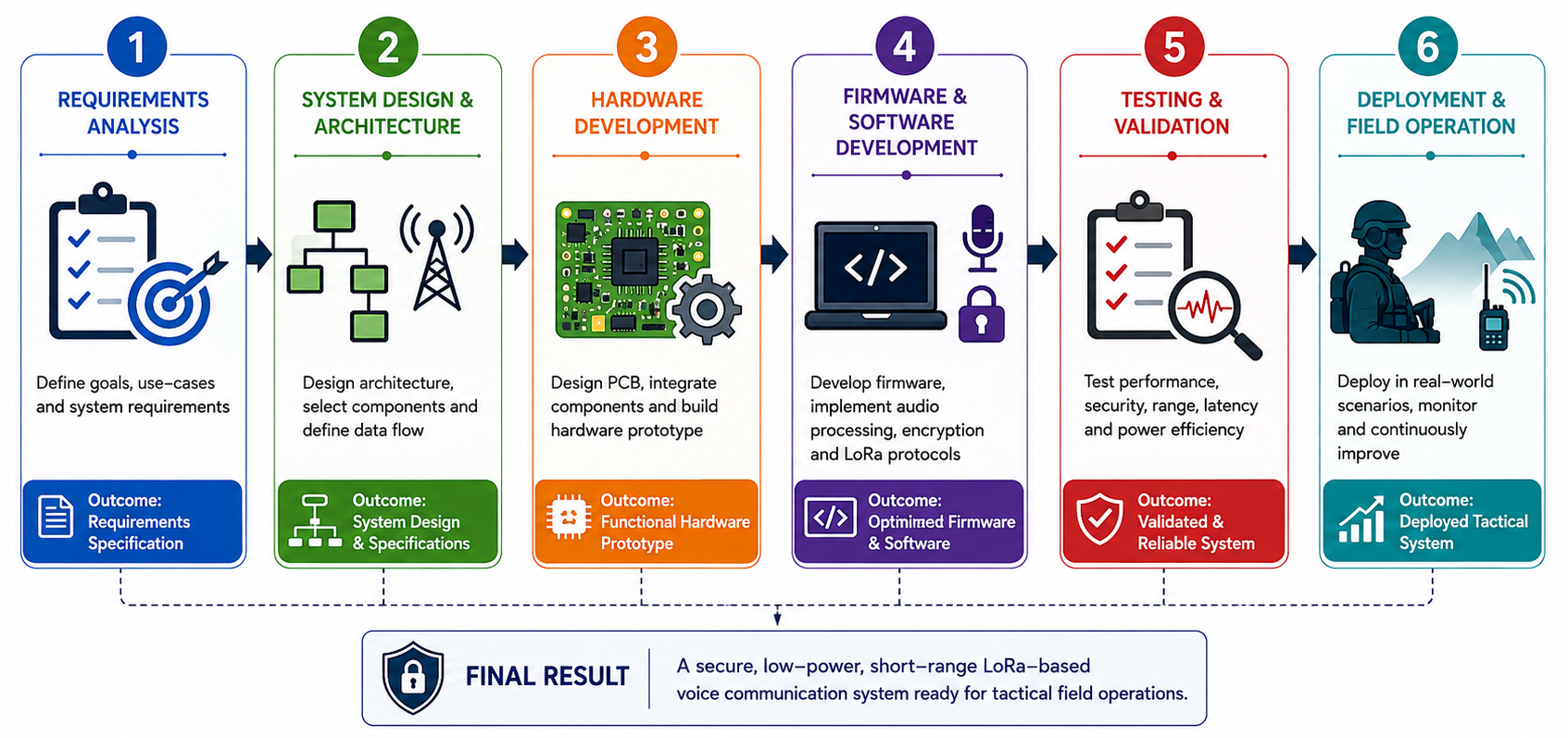}[0mm]
\Caption{Development Flow.}\flabel{5}
\end{figure}

\subsection{Architecture Design}\label{s:5.1}
The first step consisted in specifying the system-level architecture according the operational requirements including the range, latency, encrypted strength and battery life \cite{ref15}. It became a modular layered model where voice acquisition, processing, encryption and communication became separate.

Theoretical Feasibility Study: A feasibility study was conducted that involved:
\begin{itemize}
\item Link budget estimation
\item Data rate and latency analysis
\item Power consumption modeling
\item Encryption overhead assessment
\end{itemize}
The choice of component was done by taking in considerations the computational ability, power efficiency and RF ability that resulted in the choice of ESP32 and SX1276 modules.

\subsection{PCB Development}\label{s:5.2}
Subsequently, architectural validation was done followed by the schematic and PCB layout. The layout design was done with particular attention to provide consistent RF operation and a low noises operation. Key design factors were controlling the RF trace impedance, a good ground plane partitioning to ensure noise suppression, analog/digital isolation and optimization of antenna location to prevent any coupling effects.

Signal integrity analysis was carried out to reduce the electromagnetic interference between the audio codec and the RF transceiver parts. The adoption of a multilayer PCB stack-up was aimed at accomplishing small integration and minimizing parasitics and unwanted coupling paths.

\subsection{Firmware Development}\label{s:5.3}
Software development was also developed along with PCB manufacturing to reduce design cycle. The firmware architecture was modular in nature and functional blocks could be independently validated and refined. The modules that were implemented are the VOX based threshold detection, audio sampling and compression, AES-128 encryption integration, LoRa packet framing and CRC validation and low-power sleep scheduling.

Scheduling of real time lightweight tasks was used to ensure that with voice transmission, there was deterministic latency. Moreover, the encryption algorithms have been made optimization friendly so that they do not consume a lot of computer time at the expense of the security level needed.

\subsection{System Integration}\label{s:5.4}
The hardware and firmware subsystems were added gradually after being tested separately. This had the benefit of isolating faults in a staged manner and minimized the time spent in debugging when bringing the system up. Integration testing was centered around testing the continuity of audio signals, consistency in encryption and decryption, and behavior of packet loss and end to end latency.

Accuracy of modulation, timing synchronization and protocol-level correctness were confirmed by use of logic analyzers and RF spectrum analyzers when operating the system.

\subsection{Laboratory Testing}\label{s:5.5}
Indoor experiment was also carried out to test the core system functionality before it could be deployed in the field. The test was conducted based on functional check, power consumption test at varied duty cycles, voice intelligibility test at various spreading factors and packet error rate test in controlled environment of interference test.

Latency was measured by comparing the timestamps of transmission and the playback timestamps at the receiver.

\subsection{Field Trials}\label{s:5.6}
Indoor experiment was also carried out to test the core system functionality before it could be deployed in the field. The test was conducted based on functional check, power consumption test at varied duty cycles, voice intelligibility test at various spreading factors and packet error rate test in controlled environment of interference test.

Latency was measured by looking at transmission timestamps and comparing them to according playback timestamps at the receiver.

\section{Expected Outcomes}\label{s:6}
The successful implementation of the proposed system is expected to result in the provision of a compact indigenous and secure short-range tactical communication platform that is independent of the conventional communication infrastructure. By combining chirp spread spectrum modulation using LoRa with embedded AES-128 encryption, a confidential peer-to-peer voice communication framework is provided for deployment in the field, for example in defense and disaster response situations.

From the hardware point of view, the device is expected to provide a substantial reduction in size and power consumption, in comparison with conventional systems for software-defined radio (SDR) systems and legacy VHF/UHF tactical radios. The low mass configuration of the wearable improves mobility of operations and keeps the communication reliable within a line of sight distance of 1--1.5 km. A voice activated transmission integration also further optimizes energy utilization and reduced electromagnetic exposure as well by limiting unnecessary radio activity.

At the network level, the architecture provides a scalable squad-level communication model which can be expanded to the multi-node configurations with a minimum of infrastructure dependency. The encrypted packetized communication mechanism guarantees data confidentiality, integrity and resistance to unauthorized interception.

Overall, the proposed framework shows the feasibility of extending the use of the low-power wide-area communication technologies beyond traditional telemetry applications for IoT by extending its applicability toward secure, field-ready, real-time tactical voice communications systems.

\section{Conclusion}\label{s:7}
This paper presented the design and theoretical validation of a wearable LoRa-based secure intercommunication system for short-range tactical applications. By integrating voice-activated acquisition, embedded signal processing, AES-128 encryption, and chirp spread spectrum modulation, the proposed architecture enables secure peer-to-peer voice communication without relying on existing infrastructure. Analytical link budget evaluation confirms reliable communication over a 1--1.5 km line-of-sight range with substantial link margin, while latency analysis demonstrates end-to-end delays within acceptable conversational limits. The use of half-duplex transmission and adaptive duty cycling significantly improves energy efficiency, enabling 12--16 hours of continuous operation in a compact wearable form factor.

Compared with conventional VHF/UHF and SDR-based systems, the proposed framework offers lower hardware complexity, reduced power consumption, and enhanced portability. The results demonstrate the feasibility of extending LoRa technology beyond traditional IoT telemetry toward secure real-time tactical communication systems.

% Author notes are generated from the author blocks above.
\def\EndNotes{}
\wyswietlnotke{1}
\wyswietlnotke{2}
\wyswietlnotke{3}
\wyswietlnotke{4}
\wyswietlnotke{5}

\end{document}